\documentclass[twocolumn]{article}
\usepackage{graphicx}
\usepackage{amsmath,amssymb}
\usepackage{cite}
\usepackage{url}
\usepackage{algorithmic}
\usepackage{textcomp}
\usepackage{xcolor}
\usepackage{listings}
\usepackage{placeins}
\usepackage{geometry}
\usepackage{hyperref}
\lstdefinestyle{solidity}{
    language=Java,
    basicstyle=\ttfamily\tiny,
    keywordstyle=\bfseries\color{blue},
    commentstyle=\color{green!60!black},
    stringstyle=\color{red!60!black},
    numbers=left,
    numberstyle=\tiny,
    numbersep=5pt,
    frame=single,
    breaklines=true,
    captionpos=b,
    morekeywords={pragma, solidity, contract, address, uint256, function, mapping, event, emit, require},
}
\title{SmartLLM: Smart Contract Auditing using Custom Generative AI}

\author{Jun Kevin \\
Universitas Pelita Harapan \\
Jakarta, Indonesia \\
0009-0004-1304-6398 \\
\and
Pujianto Yugopuspito \\
Universitas Pelita Harapan \\
Jakarta, Indonesia \\
0000-0001-6155-163X}

\begin{document}
\maketitle

\begin{abstract}
Smart contracts, integral to decentralized finance (DeFi) and blockchain ecosystems, are increasingly vulnerable to exploits due to coding errors and complex attack vectors. Traditional static analysis tools and existing vulnerability detection methods often fail to address these challenges comprehensively, resulting in high false-positive rates and an inability to detect dynamic vulnerabilities. This paper introduces \textbf{SmartLLM}, a novel approach leveraging fine-tuned \textbf{LLaMA 3.1} models with \textbf{Retrieval-Augmented Generation (RAG)} to enhance the accuracy and efficiency of smart contract auditing. By integrating domain-specific knowledge from ERC standards and employing advanced techniques such as QLoRA for efficient fine-tuning, SmartLLM achieves superior performance compared to static analysis tools like Mythril and Slither and zero-shot LLM prompting (e.g., ChatGPT-3.5 and GPT-4). Experimental results demonstrate a perfect recall of 100\% and an accuracy score of 70.0\%, underscoring the model’s robustness in identifying vulnerabilities, including re-entrancy and access control issues. This research advances smart contract security by offering a scalable and effective auditing solution, supporting the secure adoption of decentralized applications.
\end{abstract}

\section{Introduction}
Since the inception of Bitcoin, blockchain technology has significantly evolved, with Ethereum emerging as a pivotal development. Ethereum, a decentralized and open-source blockchain platform, enables the creation and execution of decentralized applications (DApps) and smart contracts. Ethereum Request for Comments (ERCs) have been developed to standardize smart contracts \cite{ethereum}. Among these, ERC-20 is notable for defining rules for fungible tokens and ensuring interoperability across various Ethereum-based projects \cite{vogelsteller2015erc20}. However, smart contracts are not immune to vulnerabilities, and violations of ERC standards can lead to severe interoperability issues, security risks, and financial losses \cite{zhou2023sok,zhang2023demystifying}.

The rapid growth of blockchain and decentralized finance (DeFi) has transformed the financial landscape by offering new opportunities for peer-to-peer transactions and automated services. Despite their advantages, smart contracts are susceptible to attacks and vulnerabilities, including integer overflow, re-entrancy, and access control issues, which can result in significant financial damage and erode trust in the DeFi ecosystem \cite{sendner2023vulnerability,chen2023chatgpt}. As smart contracts become increasingly integral to DeFi applications, the need for robust security measures and innovative auditing techniques is crucial to mitigate these risks and ensure the security and reliability of decentralized financial systems \cite{li2022audit}.

Smart contract vulnerabilities often stem from coding flaws, significantly compromising the security and functionality of decentralized applications. Common issues include integer overflow, re-entrancy attacks, and inadequate access control mechanisms, all arising from code logic errors, improper use of programming constructs, or missing validation checks \cite{chen2023chatgpt,ghaleb2023achecker}. For instance, re-entrancy vulnerabilities occur when improper handling of external function calls allows attackers to repeatedly withdraw funds before the contract updates its balance \cite{he2020smart}. Similarly, integer overflow happens when numerical calculations exceed the storage capacity, leading to incorrect results or exploitable conditions. Access control issues pose risks where insufficient authorization checks enable unauthorized users to manipulate contract functions \cite{sendner2023vulnerability}. Given the immutable nature of blockchain, once a smart contract is deployed, its code cannot be modified, making these vulnerabilities particularly critical. This immutability means that any security flaws can be exploited without retroactive fixes, potentially leading to substantial financial losses and eroding trust in the blockchain ecosystem \cite{chen2023chatgpt,li2022audit}.

Static code analysis is a common technique for identifying vulnerabilities in smart contracts before deployment, as it examines the code without execution to detect potential risks. However, this approach has significant limitations, often struggling to identify complex vulnerabilities that depend on dynamic execution contexts, such as interactions between multiple contracts or external data sources \cite{sendner2023vulnerability,ghaleb2023achecker}. Additionally, static analysis may generate numerous false positives, requiring extensive manual verification to filter out benign code patterns \cite{chen2023chatgpt}. The approach also falls short when dealing with evolving coding practices and emerging vulnerabilities, as many tools rely on predefined patterns that may not account for unconventional attack vectors. Consequently, while static analysis is an important part of the auditing process, it alone is insufficient for ensuring the security of smart contracts, highlighting the need for more advanced techniques \cite{li2022audit}.

Recent research has explored the use of Large Language Models (LLMs) for smart contract auditing, leveraging their ability to understand and generate code to detect vulnerabilities that static analysis tools may overlook \cite{yu2024rag}. LLMs can analyze code contextually, consider dynamic scenarios, and recognize complex patterns. With techniques like Retrieval-Augmented Generation (RAG) and fine-tuning, LLMs can be adapted to the rapidly changing landscape of smart contract security \cite{li2022audit}. Integrating LLMs into the auditing process not only enhances vulnerability detection but also reduces manual verification efforts by effectively filtering false positives \cite{he2020smart}.

To address the gaps in current smart contract auditing methods, this research seeks to answer the following questions:

\begin{itemize}
    \item \textbf{RQ1}: How effective is our model in detecting specific types of vulnerabilities in smart contracts based on accuracy, precision, recall and F1 score?
    \item \textbf{RQ2}: How does our model's performance compare to existing smart contract vulnerability detection tools such as Mythril or Slither and zero-shot learning LLM such as ChatGPT-3.5 or ChatGPT-4?
    \item \textbf{RQ3}: What are the main opportunities for improving our model, and what challenges does it face in detecting complex vulnerabilities in smart contracts?
\end{itemize}

This paper contributes to the ongoing research by introducing a novel approach that integrates RAG with the documentation of ERC standards, utilizing the latest LLaMA 3.1 8B model. The proposed method introduces a new role, the \textit{verificator}, responsible for fact-checking vulnerabilities using RAG with ERC documentation, alongside a reasoning role. This approach aims to enhance the accuracy of smart contract vulnerability detection, offering insights into the strengths and limitations of LLM-based auditing and providing recommendations for future advancements in the field.

\section{Literature Review}

\subsection{ERC-20}
The ERC-20 standard defines a set of rules for fungible tokens on the Ethereum blockchain, enabling interoperability between decentralized applications (DApps) and ensuring uniformity in token behavior \cite{ansari2020erc20}. The standard specifies six key functions, including \textit{transfer} for token transfers and \textit{approve} for delegated spending, along with two events, \textit{Transfer} and \textit{Approval}, to log token transactions and approvals \cite{durieux2020icse}.

Despite its widespread adoption, ERC-20 compliance is not without challenges. Violations of the standard can lead to issues such as unhandled return values in \textit{transfer} functions or insufficient validation of input parameters, which increase the risk of vulnerabilities \cite{ansari2020erc20}. These challenges underline the need for automated auditing tools to ensure compliance and security in ERC-20 implementations.

\subsection{Static Code Analysis}
Static code analysis is a prevalent method for identifying vulnerabilities in smart contracts before deployment. It analyzes code without executing it, allowing tools like Mythril and Slither to detect common issues such as reentrancy and integer overflow \cite{durieux2020icse, feist2019slither}.

However, static analysis has significant limitations. It struggles to detect vulnerabilities arising from dynamic execution contexts, such as those involving external contract interactions or real-time data dependencies \cite{durieux2020icse}. Additionally, the reliance on predefined patterns often leads to high false positive rates, requiring extensive manual verification to filter benign code patterns. These limitations highlight the need for more advanced auditing methods, such as those leveraging generative AI.

\subsection{Generative AI}
Generative AI represents a class of artificial intelligence systems designed to generate coherent, contextually relevant outputs based on input data \cite{feuerriegel2024generative}. These models are built on large-scale neural networks, such as transformers, which excel at understanding patterns in text and generating human-like responses. The rise of Generative AI has transformed various domains, including natural language processing, code generation, and automated reasoning \cite{brynjolfsson2023generative}.

In the context of smart contract auditing, Generative AI models, such as GPT and LLaMA, offer a significant advantage over traditional static analysis tools \cite{hu2024llm}. By leveraging contextual understanding and pattern recognition, these models can detect vulnerabilities that are often missed by rule-based methods. For instance, Generative AI can analyze reentrancy vulnerabilities by understanding both the code and its dynamic execution context, enabling more comprehensive assessments \cite{chen2023chatgpt}.

Generative AI also introduces flexibility through techniques like fine-tuning and retrieval-augmented generation (RAG) \cite{ding2023parameter, lewis2020rag}. These capabilities allow models to adapt to specific domains, such as smart contract security, by incorporating domain-specific knowledge bases like ERC documentation. However, challenges remain, including the computational cost of training and inference, reliance on large annotated datasets, and the need for interpretability in critical applications.

Despite these challenges, the application of Generative AI in vulnerability detection is a promising frontier, bridging the gap between static analysis and dynamic reasoning to enhance the security of decentralized systems.

\subsection{LLaMA 3.1}
LLaMA (Large Language Model Meta AI) is a series of large language models designed to deliver state-of-the-art performance with optimized computational efficiency \cite{touvron2023llama}. The LLaMA 3.1 model introduces enhancements in contextual understanding and computational efficiency, making it particularly suited for tasks requiring deep code analysis and reasoning.

In the context of smart contract auditing, LLaMA 3.1's ability to process long sequences and recognize complex patterns in code positions it as a valuable tool. By integrating this model with fine-tuning techniques and domain-specific knowledge bases, it can effectively identify vulnerabilities that static analysis tools may overlook.

\subsection{QLoRA: Efficient Finetuning of Quantized LLMs}
QLoRA (Quantized Low-Rank Adaptation) is an efficient fine-tuning technique that significantly reduces the memory and computational requirements for large language models \cite{dettmers2023qlora}. By employing low-rank adaptation with 4-bit quantization, QLoRA enables the fine-tuning of models like LLaMA 3.1 on resource-constrained hardware without compromising performance.

\begin{figure}[h!]
    \centering
    \includegraphics[width=0.5\linewidth]{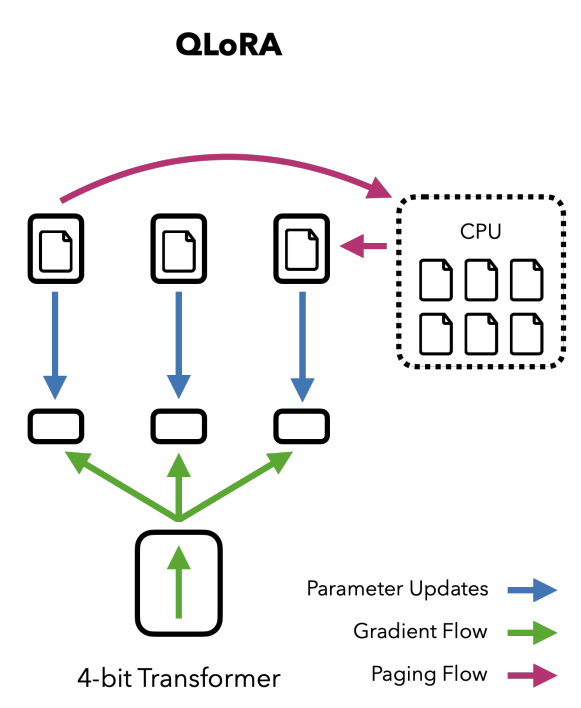}
    \caption{How QLORA quantizing the model to 4-bit precision and using paged optimizers to handle memory spikes \cite{dettmers2023qlora}.}
    \label{fig:qlora}
\end{figure}

This technique uses paged optimizers to handle memory spikes efficiently, as illustrated in Fig.~\ref{fig:qlora}, allowing for smoother training on GPUs with limited memory capacity. In this research, QLoRA is applied to fine-tune LLaMA 3.1 using domain-specific datasets, such as smart contract code and ERC documentation. This approach ensures the model is adapted efficiently for smart contract vulnerability detection while maintaining feasibility for deployment on standard hardware setups.

\section{Research Methodology}

\subsection{Data Collection}
The dataset for this research includes smart contract code sourced from Etherscan, a widely-used platform for Ethereum blockchain data. The dataset includes labeled examples of vulnerable and non-vulnerable contracts, which were essential for training and evaluation. The dataset includes 300 smart contract samples, equally split between vulnerable and non-vulnerable contracts:
\begin{itemize}
    \item \textbf{Vulnerable contracts:} 150 samples with three distinct vulnerability types:
    \begin{itemize}
        \item \textbf{Re-Entrancy:} Improper handling of external calls, allowing repeated execution (\textbf{49 samples}).
        \item \textbf{Access Control:} Insufficient authorization checks (\textbf{40 samples}).
        \item \textbf{Logic Error:} Errors in code logic leading to unintended behaviors (\textbf{61 samples}).
    \end{itemize}
    \item \textbf{Non-vulnerable contracts:} 150 samples with correct implementations.
\end{itemize}

The dataset was split into training and testing subsets using 80:20 ratio. Specifically, 240 contracts (120 vulnerable and 120 non-vulnerable) were used for training, while the remaining 60 contracts (30 vulnerable and 30 non-vulnerable) were used for testing.

To illustrate the vulnerabilities present in the dataset, a simplified example of a vulnerable ERC-20 smart contract is shown in Listing~\ref{lst:vulnerable-erc20-short}. The corpus for RAG comprises ERC-20 documentation \cite{ethereum, vogelsteller2015erc20}.

\begin{lstlisting}[caption={Sample of Vurnerability Code}, label={lst:vulnerable-erc20-short}]
pragma solidity ^0.8.0;

contract ERC20 {
    string public name = "Token101";
    string public symbol = "TKN";
    uint256 public totalSupply = 1000000;

    mapping(address => uint256) public balanceOf;
    mapping(address => mapping(address => uint256)) public allowance;

    constructor() {
        balanceOf[msg.sender] = totalSupply;
    }

    function transfer(address _to, uint256 _value) public returns (bool) {
        balanceOf[msg.sender] -= _value; // No check for sufficient balance
        balanceOf[_to] += _value;
        return true;
    }

    function transferFrom(address _from, address _to, uint256 _value) public returns (bool) {
        balanceOf[_from] -= _value; // Missing allowance check
        balanceOf[_to] += _value;
        (bool success, ) = _to.call{value: _value}(""); // Unsafe external call
        require(success, "Transfer failed");
        return true;
    }
}
\end{lstlisting}

\subsection{Data Preprocessing}
Prior to training, the smart contract code was preprocessed to remove irrelevant comments, normalize formatting, and tokenize using the LLaMA tokenizer. The corpus for RAG was indexed using FAISS, enabling efficient retrieval of relevant context during the inference phase \cite{douze2024faiss}. This preprocessing ensured consistency and optimized the model's performance.

\subsection{Model Fine-Tuning with QLoRA}
QLoRA (Quantized Low-Rank Adaptation) was employed to fine-tune the LLaMA 3.1 model on the prepared dataset. This technique reduces memory requirements by applying 4-bit quantization and low-rank adaptation, allowing efficient training on GPUs with limited resources \cite{dettmers2023qlora}. Training parameters included a learning rate of $1e^{-4}$, a batch size of 16, and a training duration of 2 epochs.

\subsection{Instruction Scenarios}
\begin{figure}[h!]
    \centering
    \includegraphics[width=0.5\textwidth]{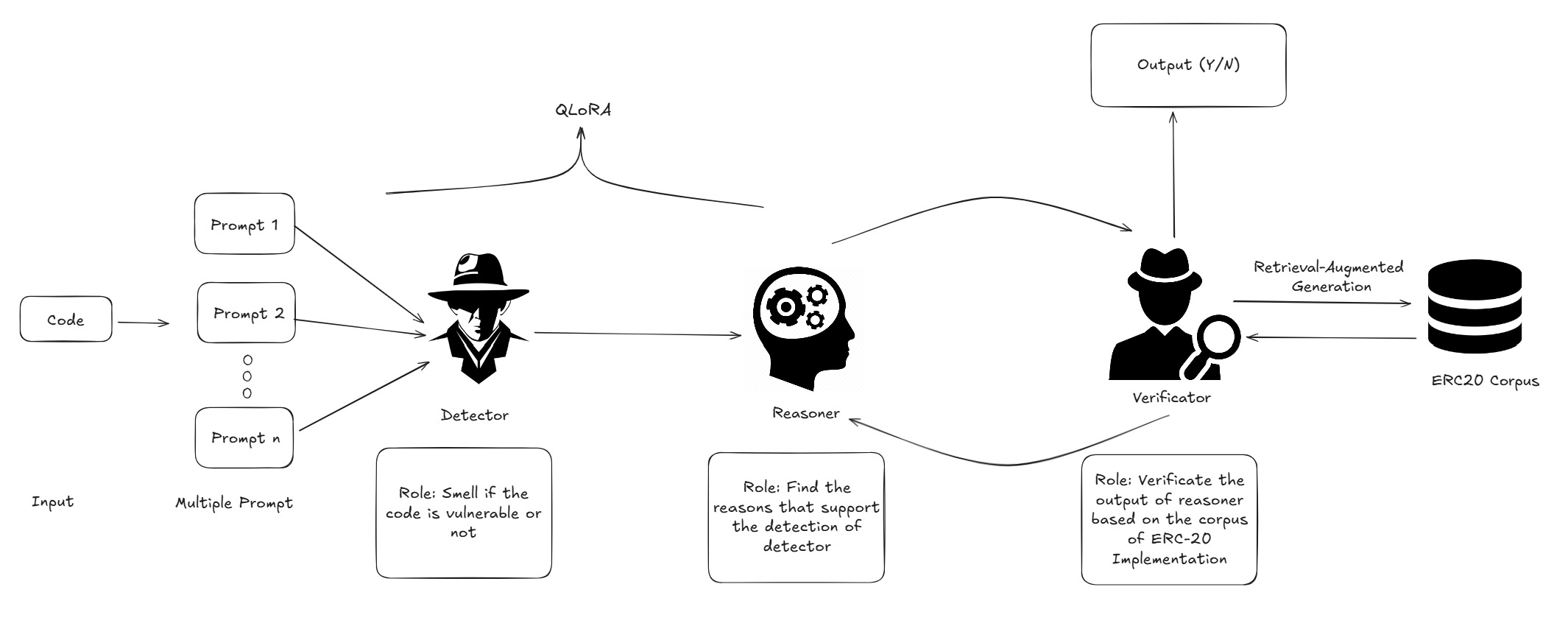}
    \caption{Workflow diagram of SmartLLM, illustrating the roles of Detector, Reasoner, and Verificator in vulnerability detection using Retrieval-Augmented Generation.}
    \label{fig:workflow_diagram}
\end{figure}

To evaluate the model's ability to understand and detect vulnerabilities, three distinct roles were implemented within the workflow, as shown in Fig.~\ref{fig:workflow_diagram}:

\begin{enumerate}
    \item \textbf{Detector:} The Detector, illustrated in Fig.~\ref{fig:workflow_diagram}, analyzes multiple prompts to classify smart contracts as vulnerable (\texttt{Y}) or non-vulnerable (\texttt{N}). This role ensures that potential vulnerabilities in the code are flagged for further analysis.
    \item \textbf{Reasoner:} Following the Detector's output, the Reasoner identifies and provides reasons supporting the classification decision. As depicted in the workflow, this role offers detailed explanations for the detected vulnerabilities.
    \item \textbf{Verificator:} The Verificator validates the outputs of the Reasoner using Retrieval-Augmented Generation (RAG) to cross-reference findings against ERC-20 corpus. This role, also illustrated in Fig.~\ref{fig:workflow_diagram}, ensures compliance and accuracy before generating the final classification.
\end{enumerate}

These instruction scenarios were designed to comprehensively test the model's binary classification capability, its reasoning for nuanced vulnerability detection, and its ability to validate results against domain-specific knowledge.

\subsection{Evaluation Metrics}
The effectiveness of the proposed approach was evaluated using the following metrics:
\begin{itemize}
    \item \textbf{Confusion Matrix:} A table summarizing the model performance by showing the counts of true positives (TP), true negatives (TN), false positives (FP) and false negatives (FN). This provides an overview of the model's classification capabilities.
    \item \textbf{Accuracy:} The proportion of correct predictions across all samples, calculated as:
    \begin{equation}
        \text{Accuracy} = \frac{\text{TP} + \text{TN}}{\text{TP} + \text{TN} + \text{FP} + \text{FN}}
    \end{equation}
    \item \textbf{Recall:} The model's ability to identify all vulnerable contracts, calculated as:
    \begin{equation}
        \text{Recall} = \frac{\text{TP}}{\text{TP} + \text{FN}}
    \end{equation}
    \item \textbf{F1 Score:} The harmonic mean of precision and recall, providing a balanced measure of the model's accuracy and completeness:
    \begin{equation}
        \text{F1 Score} = 2 \cdot \frac{\text{Precision} \cdot \text{Recall}}{\text{Precision} + \text{Recall}}
    \end{equation}
    \item \textbf{Precision:} The proportion of correctly predicted vulnerabilities out of all positive predictions, calculated as:
    \begin{equation}
        \text{Precision} = \frac{\text{TP}}{\text{TP} + \text{FP}}
    \end{equation}
\end{itemize}

These metrics provide a comprehensive evaluation of the model's performance, enabling comparisons with static analysis tools (e.g., Mythril, Slither) and zero-shot prompting with large language models.

\section{Results and Discussion}
\subsection{Experimental Setup}
To evaluate the proposed methodology, we compared its performance against two categories of benchmarks: static code analysis tools and zero-shot prompting with large language models (LLM). For static code analysis, we selected industry-standard tools, including Mythril and Slither. For zero-shot prompting, we utilized advanced LLMs, including GPT-3.5 and GPT-4. These models were assessed without any task-specific fine-tuning, relying solely on their pre-trained knowledge to classify smart contracts as vulnerable (\texttt{Y}) or non-vulnerable (\texttt{N}). This setup tested the LLMs' ability to reason over smart contract code and provide accurate classifications directly.

Our proposed method differs significantly from these benchmarks. By integrating Retrieval-Augmented Generation (RAG), the model retrieves domain-specific knowledge from the ERC documentation during classification, enhancing its contextual understanding. Unlike static code analysis tools or zero-shot prompting, the model generates label names directly as outputs, improving interpretability and adaptability to complex scenarios.

\subsection{Results}

\begin{figure}[h!]
    \centering
    \includegraphics[width=0.7\linewidth]{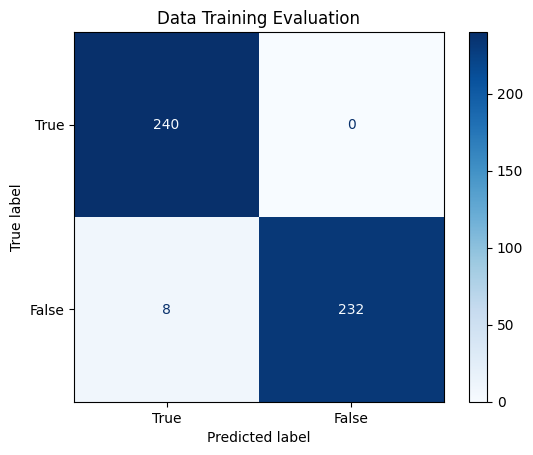}
    \caption{Confusion Matrix for Training Dataset}
    \label{fig:confusion-matrix-training}
\end{figure}

\begin{figure}[h!]
    \centering
    \includegraphics[width=0.7\linewidth]{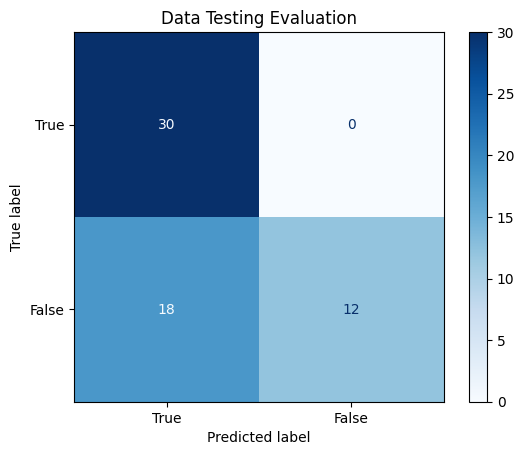}
    \caption{Confusion Matrix for Testing Dataset}
    \label{fig:confusion-matrix-testing}
\end{figure}

The evaluation of the proposed model was carried out on both training and testing datasets. Table~\ref{tab:metrics} summarizes the performance metrics, including accuracy, recall, precision, and the F1 score. In addition, Figure~\ref{fig:confusion-matrix-training} and Figure~\ref{fig:confusion-matrix-testing} illustrate the confusion matrices for the training and testing datasets, respectively, providing a detailed breakdown of true positives, false positives, true negatives, and false negatives. 

\begin{table}[h!]
\centering
\caption{Evaluation Metrics for Training and Testing Data}
\label{tab:metrics}
\begin{tabular}{|l|c|c|}
\hline
\textbf{Metric}      & \textbf{Training Data} & \textbf{Testing Data}            \\ \hline
Accuracy             & 98.3\%                & 70.0\%                           \\ \hline
Recall               & 100\%              & 100\%                          \\ \hline
Precision            & 96.8\%                & 62.5\%                           \\ \hline
F1 Score             & 98.4\%                & 76.9\%                           \\ \hline
\end{tabular}
\end{table}

The model demonstrated exceptional learning capabilities during training, achieving near-perfect metrics. In the test data set, the model maintained perfect recall, ensuring that all vulnerabilities were identified, with F1 score of 76.90\%. The perfect recall (100\%) ensures that no vulnerabilities are overlooked, making the model reliable for critical applications. However, the trade-off with accuracy (70.0\%) and precision (62.5\%) suggests the need for further fine-tuning to reduce false positives, which is vital to minimizing manual verification efforts.

These results highlight the model’s robustness in identifying specific types of vulnerabilities, such as re-entrancy and access control issues. However, the slightly lower precision in the testing phase suggests room for improvement in reducing false positives. Overall, the model is highly effective and reliable for smart contract vulnerability detection.

Table~\ref{tab:comparison} presents a comprehensive comparison of vulnerability detection methods, including static analysis tools (Mythril and Slither), zero-shot prompting with large language models (ChatGPT-3.5 and ChatGPT-4), and the proposed approach (SmartLLM).

The static analysis tools, Mythril and Slither, were run locally on a machine with 32GB of RAM and an Intel Core Ultra 7 processor, ensuring consistent performance for comparison. For SmartLLM, the model was fine-tuned and evaluated using Google Colab Pro with NVIDIA A100 GPU and 40GB of VRAM, leveraging the platform's high computational capacity for efficient processing of large datasets.

\begin{table}[h!]
\centering
\caption{Comparison of Vulnerability Detection Methods}
\label{tab:comparison}
\begin{tabular}{|l|c|c|c|c|}
\hline
\textbf{Method} & \textbf{Accuracy} & \textbf{Recall} & \textbf{Precision} & \textbf{F1 Score} \\ \hline
Mythril         & 21.7\%          & 34.8\%         & 20.0\%            & 25.4\%           \\ \hline
Slither         & 35\%          & 42.3\%         & 31.4\%            & 36.1\%           \\ \hline
ChatGPT-3.5     & 39.1\% & 38.7\%        & 37.5\%            & 38.1\%           \\ \hline
ChatGPT-4       & 43.3\% & 36.7\%         & 42.3\%            & 39.3\%           \\ \hline
\textbf{SmartLLM} & \textbf{70.0\%} & \textbf{100\%} & \textbf{62.5\%} & \textbf{76.9\%} \\ \hline
\end{tabular}
\end{table}

Traditional static analysis tools, Mythril and Slither, show limited effectiveness in identifying vulnerabilities. Mythril achieved an accuracy of 21.7\%, with recall and precision scores of 34.8\% and 20\%, respectively. These results indicate that Mythril struggles to balance false positives and false negatives, resulting F1 score of 25.4\%. Slither outperformed Mythril slightly, achieving an accuracy of 35\%, a recall of 42.3\%, and a precision of 31.4\%. However, the F1 score of 36.1\% suggests room for improvement, particularly in precision.

Zero-shot prompting with LLMs (ChatGPT-3.5 and ChatGPT-4) provided moderate performance. ChatGPT-3.5 achieved an accuracy of 39.1\%, recall of 38.7\%, precision of 37.5\%, and F1 score of 38.1\%. ChatGPT-4 demonstrated slightly better precision at 42.3\% and F1 score of 39.3\%, despite having lower recall at 36.7\%. These results highlight the potential of LLMs in vulnerability detection tasks but also underline the limitations of zero-shot prompting in complex scenarios, such as handling re-entrancy vulnerabilities or access control issues.

The proposed approach, SmartLLM, significantly outperformed all baseline methods. It achieved the highest accuracy (70.0\%), precision (62.5\%), and F1 score (76.9\%), with a perfect recall of 100\%. These results indicate that the model excels at identifying all true vulnerabilities while maintaining a balanced performance across metrics. The high recall demonstrates the reliability of the model in detecting various types of vulnerabilities, including re-entrancy and logic errors, while the improved precision reflects its ability to minimize false positives compared to traditional tools and zero-shot LLMs.

Overall, the findings emphasize the effectiveness of the SmartLLM approach, particularly in scenarios where detecting critical vulnerabilities is paramount. While static tools and zero-shot prompting provide baseline capabilities, the integration of domain-specific knowledge and fine-tuning in SmartLLM delivers superior results, making it a robust solution for smart contract vulnerability detection.

\subsection{Discussion}

Table~\ref{tab:comparison} highlights the strengths and limitations of the proposed SmartLLM approach in detecting vulnerabilities. While achieving outstanding recall (100\%), a strong F1 score (76.9\%) and accuracy (70.0\%), there are key opportunities for improvement and challenges to address.

\textbf{Opportunities:} Improving precision (currently 62.5\%) is essential to reduce false positives, which can be achieved through fine-tuning with more diverse and balanced datasets. Additionally, integrating dynamic execution contexts, expanding datasets to cover emerging vulnerabilities, and leveraging multi-modal inputs (e.g., audit reports) present significant opportunities to enhance the model's capability.

\textbf{Challenges:} The model faces token limitations that hinder processing long contracts, difficulties in generalizing across platforms, and interpreting complex attack vectors like multi-step vulnerabilities. Furthermore, the computational cost of fine-tuning and deploying large models remains a practical challenge for real-time applications.

\section{Conclusion}
This study introduced \textbf{SmartLLM}, a novel approach leveraging \textbf{Retrieval-Augmented Generation (RAG)} and fine-tuned \textbf{LLaMA 3.1} models for smart contract vulnerability detection, with a focus on improving accuracy, recall, precision, and F1 scores. By integrating ERC documentation into the vulnerability detection process, the model addresses gaps in static analysis tools and zero-shot large language models (LLMs), delivering superior results.

The proposed model achieved significant advancements over traditional tools like Mythril and Slither, and over zero-shot LLM prompting (e.g., ChatGPT-3.5 and ChatGPT-4). SmartLLM demonstrated a high recall of 100\%, ensuring comprehensive detection of vulnerabilities while achieving a balanced precision of 62.5\% and a strong F1 score of 76.9\% and accuracy 70.0\%. These results indicate the model’s robustness in detecting common and complex vulnerabilities, such as re-entrancy and access control issues, surpassing conventional methods.

Key findings highlight the effectiveness of domain-specific knowledge integration and fine-tuning techniques like QLoRA in enhancing detection accuracy. The proposed system successfully bridges the gap between static analysis tools and dynamic reasoning capabilities, providing a reliable and scalable solution for auditing Ethereum smart contracts.

Despite these achievements, challenges remain, including improving the precision to reduce false positives and addressing computational resource constraints during model fine-tuning and deployment. Future work should focus on incorporating dynamic execution contexts, broader datasets covering emerging attack vectors, and larger pre-trained models (e.g., LLaMA 70B and 405B) to enhance both precision and scalability.

In summary, \textbf{SmartLLM} represents a significant step forward in ensuring the security and reliability of smart contracts, contributing to the advancement of auditing techniques in decentralized finance (DeFi) and blockchain ecosystems. By enhancing vulnerability detection and reducing manual verification efforts, the proposed approach supports the broader adoption of secure and trustworthy decentralized applications.

\section*{Acknowledgment}
This research is partially funded by Center of Research and Community Development (LPPM), Universitas Pelita Harapan, \textbf{No. 155/LPPM-UPH/VII/2024} and \textbf{No. 156/LPPM-UPH/VII/2024}, as registered grants of the Faculty of Information Technology \textbf{No. P-113-FIT/VII/2024} and \textbf{P-114-FIT/VII/2024}, in July 2024.

\vspace{12pt}

\end{document}